\documentclass[twocolumn,aps,10pt,showpacs,amsmath,superscriptaddress,pra]{revtex4-2}
\usepackage{physics}
\usepackage[utf8]{inputenc}
\usepackage[latin9]{luainputenc}
\usepackage{mathtools}
\usepackage{amsmath}
\usepackage{braket}
\usepackage{breqn}
\usepackage{amssymb}
\usepackage{graphicx}
\usepackage[symbol]{footmisc}
\usepackage{bm}
\usepackage{siunitx}
\usepackage{upgreek}

\makeatletter

\usepackage{dcolumn}
\usepackage{bm}
\usepackage{ifpdf}
\usepackage{lineno}\usepackage{amsfonts}
\usepackage{bbm}
\usepackage{mathrsfs}
\usepackage{upgreek}
\usepackage{epstopdf}
\usepackage{setspace}
\usepackage[matrix,frame,arrow]{xypic}
\usepackage{rotating}
\usepackage{float}
\usepackage{multirow}
\usepackage[english]{babel}
\usepackage{hyperref}
\usepackage{color}
\definecolor{med-blue}{RGB}{25,25,112}
\hypersetup{colorlinks, linkcolor={blue},citecolor={blue}, urlcolor={blue}}

\renewcommand{\ket}[1]{\vert{#1}\rangle}

\newcommand\IITM{\,Department of Physics, Indian Institute of Technology Madras, Chennai 600036, India}
\newcommand\CQUICC{\,Center for Quantum Information, Communication and Computing, Indian Institute of Technology Madras, Chennai 600036, India}
\newcommand\LCN{\,London center for Nanotechnology, UCL, London WC1H0AH, UK}
\newcommand\EEE{\,Department of Electrical and Electronic Engineering, UCL, London WC1E7JE, UK}
\newcommand\krea{\,SIAS, Krea University, Sri City 517646, India}

\makeatother

\begin{document}


\title{Identifying optimal magnetic field configurations for decoherence mitigation of boron vacancies in hexagonal boron nitride}

\author{Basanta Mistri}
\email{basanta@smail.iitm.ac.in}
\affiliation{\IITM}
\affiliation{\CQUICC}
\author{Saksham Mahajan}
\affiliation{\EEE}
\author{Felix Donaldson}
\affiliation{\LCN}
\author{Rama K. Kamineni}
\affiliation{\krea}
\author{Siddharth Dhomkar}
\email{sdhomkar@physics.iitm.ac.in}
\affiliation{\IITM}
\affiliation{\CQUICC}
\affiliation{\LCN}

\date{\today}

\begin{abstract}
    The negatively charged boron vacancy center in 2D hexagonal boron nitride has emerged as a promising quantum sensor. However, its sensitivity is constrained due to ubiquitous nuclear spins in the environment. The nuclear spins, hyperfine coupled with the central electron spin, effectively behave as magnetic field fluctuators, leading to rapid decoherence. Here, we explore the effectiveness of static magnetic field strength and orientation in realizing peculiar subspaces that can lead to enhanced spin coherence. Specifically, using detailed numerical simulations of the spin Hamiltonian, we identify specific field configurations that minimize energy gradients and, consequently, are expected to facilitate decoherence suppression. We also develop an approximate analytical model based on the perturbation theory that accurately predicts these low-gradient subspaces for magnetic fields aligned with the electron spin quantization axis, applicable not only to boron vacancies but to any spin-1 electronic system coupled to nearby nuclear spins. Furthermore, to stimulate experimental validation, we estimate coherence lifetimes as a function of various bias field configurations and demonstrate that significant decoherence suppression can indeed be achieved in certain regions. These findings and the developed methodology offer valuable insights for mitigating decoherence in a low-field regime.
\end{abstract}
\maketitle
\section{Introduction}
Optically active spin defects are gaining significant attention due to their potential in various quantum applications \cite{Wolfowicz2021,Awschalom2018,Fang2024}, including quantum sensing \cite{{Balasubramanian2008},{Maze2008},{Kraus2014},{Simin2016},Fang2024,Alsid2023}, quantum computing \cite{Simmons2011,Taminiau2014,Cramer2016,Bradley2019,Bourassa2020}, and quantum communication \cite{Wolfowicz2015,Hensen2015}. A key feature of these defects is their extraordinary sensitivity to external perturbations such as magnetic fields \cite{Balasubramanian2008,Maze2008,Kraus2014,Simin2016,Alsid2023,Lamba2024}, electric fields \cite{Udvarhelyi2023,Dolde2011}, strain \cite{Udvarhelyi2023,Lyu2022}, and temperature \cite{Gottscholl2020,Neumann2013}. This sensitivity, coupled with the ability to perform optical readout \cite{Gottscholl2020,Stern2022} of spin states at room temperature, enables the development of ultrasensitive nanoscale sensors. Among the most extensively studied systems is the nitrogen-vacancy ($NV$) center \cite{Jelezko2006,Rondin2014,Schirhagl2014,Barry2020} in diamond and spin defects in silicon carbide (SiC) \cite{Castelletto2024,Jiang2023}, which have paved the way for practical implementations in quantum sensing and metrology.
Despite their success, spin defects in three-dimensional (3D) materials such as diamond and SiC face challenges that limit their applicability. Specifically, their confinement within the bulk structure prevents spin probes from approaching the target sample with sub-nanometer resolution. In contrast, spin defects in two-dimensional (2D) materials offer a promising solution. Their atomic-scale thickness allows closer proximity to the sample \cite{Huang2020,Reserbat-Plantey2021}, enhancing spatial resolution. Moreover, 2D defects can be seamlessly integrated with photonic nanostructures \cite{Reserbat-Plantey2021,Rao2024,Do2024,Sortino2024,Atature2018} enabling the development of compact, on-chip quantum technologies \cite{Branny2017,Palacios-Berraquero2017}.
Among the spin defects in 2D materials \cite{Fang2024}, the negatively charged boron vacancy ($V_B^-$) in hexagonal boron nitride (hBN) 
 \cite{Tran2016,Exarhos2019,Gottscholl2020,Liu2022} has recently emerged as a leading contender. hBN, a Van der Waals material, provides a robust and tunable platform for hosting spin defects. The $V_B^-$ defect has demonstrated remarkable potential due to its optical and spin properties \cite{Stern2022,Gottscholl2020}, making it a promising candidate for high-resolution quantum sensing \cite{Gottscholl2021,Healey2023,Lyu2022,Liu2021}. Both theoretical and experimental \cite{Ivady2020,Vaidya2023,daly2025prospects} studies have highlighted the potential of $V_B^-$ in applications ranging from nanoscale magnetometry to temperature sensing. 
\begin{table*}[ht!]
\centering
\renewcommand{\arraystretch}{1.2}
\begin{tabular}{|>{\centering\arraybackslash}m{1.5cm} 
                |>{\centering\arraybackslash}m{1.5cm} 
                |>{\centering\arraybackslash}m{1.5cm} 
                |>{\centering\arraybackslash}m{1.5cm} 
                |>{\centering\arraybackslash}m{1.5cm} 
                |>{\centering\arraybackslash}m{1.5cm} 
                |>{\centering\arraybackslash}m{1.5cm} 
                |>{\centering\arraybackslash}m{1.5cm} 
                |>{\centering\arraybackslash}m{1.5cm}|}
\hline
 & $A_{xx} $ & $A_{yy} $ & $A_{zz} $ & $A_{xy} $ & $Q_{xx} $ & $Q_{yy} $ & $Q_{zz} $ & $Q_{xy} $ \\
\hline
$^{14}N_1$ & 46.944 & 90.025 & 48.158 & 0 & -0.46 & 0.98 & -0.52 & 0 \\
\hline
$^{14}N_2$ & 79.406 & 58.170 & 48.159 & -18.391 & 0.62 & -0.1 & -0.52 & -0.623 \\
\hline
$^{14}N_3$ & 79.406 & 58.170 & 48.159 & 18.391 & 0.62 & -0.1 & -0.52 & 0.623 \\
\hline
\end{tabular}
\caption{The table outlines all the hyperfine and quadrupolar (in MHz) tensor elements used in the numerical calculations.}
\label{Table I}
\end{table*}
A key goal within quantum sensing research is to enhance sensitivities through increasing the coherence time $(T_2)$ of the probe system. In solid state systems the primary source of decoherence is magnetic field fluctuations in the spin bath surrounding the defect. On the materials engineering side, isotopic purification has been a successful strategy in systems like diamond \cite{Balasubramanian2009} and SiC \cite{Tyryshkin2012}, however, hBN naturally lacks this avenue for isotopic refinement because of its chemical composition \cite{Lee2022}. Therefore, alternative strategies to suppress decoherence are crucial to realize the full potential of $V_B^-$ defects in hBN.
Besides conventional dynamical decoupling methods \cite{Hahn_1950,Carr_1954}, a promising approach to address decoherence is identifying subspaces of the system's Hilbert space that are minimally affected by spin fluctuations \cite{Wolfowicz2013,Balian2015,Shin2013,McAuslan2012,Nicolas2023,Davidsson2024,RamaKoteswaraRao2020} in the bath. This requires a detailed understanding of how external parameters, such as magnetic field, influence the defect's energy level structure and spin-transition properties. Recent experimental and simulation studies \cite{tarkanyi2025understanding,scholten2025optimizing,mahajan2025impactirradiationcondition} shows that spin coherence in hBN defects is strongly governed by interactions with nearby nuclear spins and the applied magnetic field strength. 
Our work builds on and extends these recent advances by combining detailed numerical simulations of the spin Hamiltonian with an analytical perturbation theory model. Specifically, we investigate the influence of the strength and the orientation of the static magnetic field on the transition energies of $V_{B}^-$ defect in hBN. We analyze energy gradient and curvature at low-bias magnetic fields ($<$ 25 mT), identifying regions where decoherence effects can be minimized. We show that these regions with minimal sensitivity to the magnetic fluctuation correspond to highly improved coherence and are critical for a variety of quantum sensing applications. 
\section{System Hamiltonian}
\begin{figure}[h]
    \centering
    \includegraphics[width=0.45\textwidth, trim={0 2.5cm 0 0}, clip]{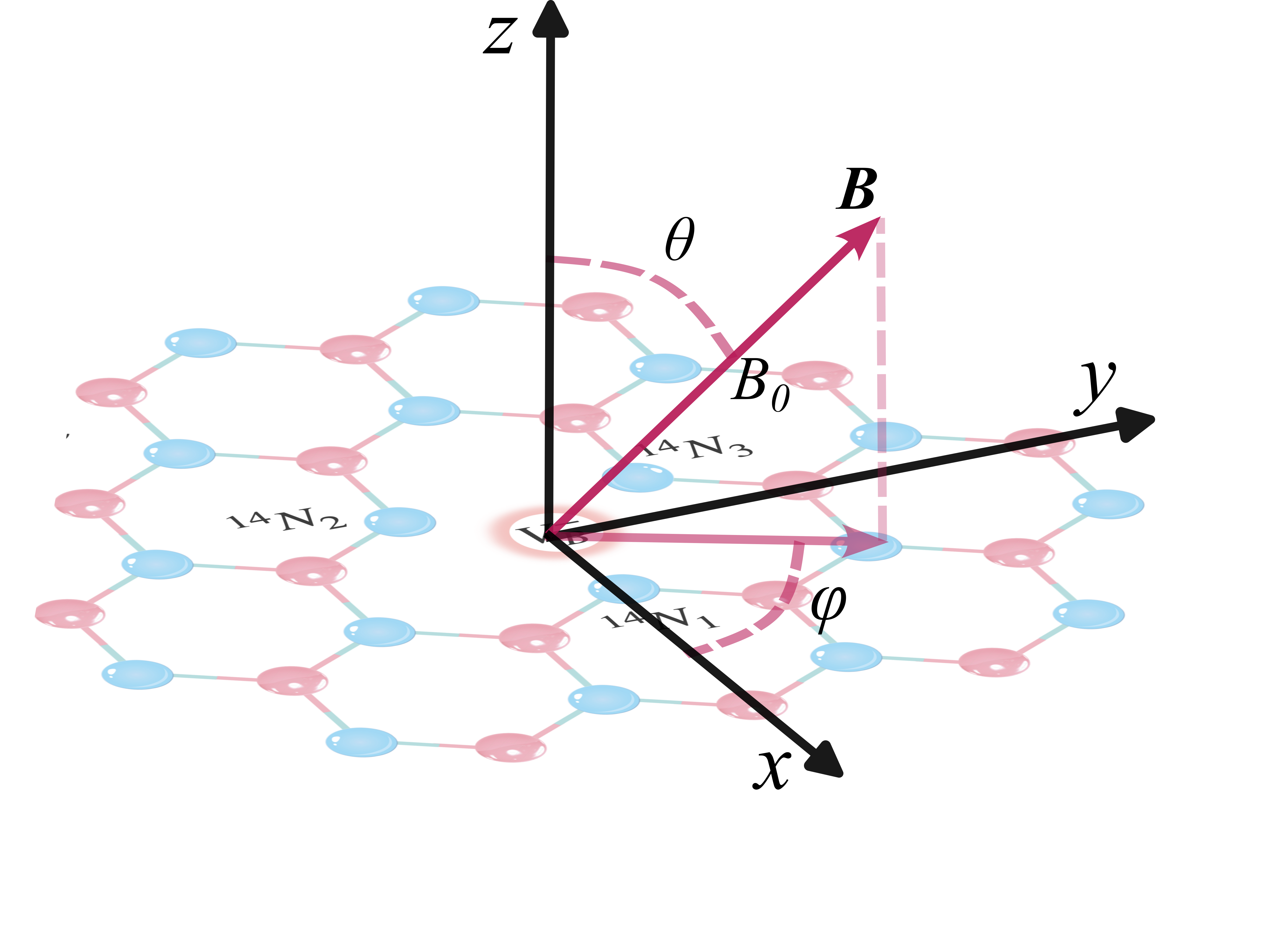}
    \caption{
    Structure of the $V_B^-$ defect in hBN, showing the three nearest-neighbor $^{14}\mathrm{N}$ nuclear spins. $\theta$ and $\varphi$ represent the polar and azimuthal angles associated with the applied magnetic field $\bm{B}$ of magnitude $B_0$. Pink and blue spheres denote boron and nitrogen atoms, respectively, while the red circle marks the boron vacancy site. The coordinate axes and magnetic field orientation are indicated for reference.}
    \label{fig:1}
\end{figure}

A missing boron atom in 2D-hBN creates a boron vacancy ($V_{B}$) defect. In its negatively charged state, this defect hosts a spin-1 electronic system, surrounded by a spin-full nuclear bath, shown in schematic Fig. \ref{fig:1}. The three $1^{st}$ shell $^{14}N$ nuclear spins (spin-1) are strongly coupled with the central $V_{B}^-$ electron spin (spin-1) via hyperfine interaction. The Hamiltonian of this composite system in the presence of an external magnetic field $\left(\bm{B}\right)$ can be written as:
\begin{dmath}
    H = D\left(S_z^2-\frac{2}{3}\right) + \epsilon \left(S_y^2 - S_x^2\right) + \gamma_e  \bm{B}.\bm{S} + \sum_{i=1}^{3} \bm{S}.\bm{A}^i.\bm{I}^i +  \sum_{i=1}^{3} \gamma_{n}^i \bm{B}.\bm{I}^i + \sum_{i=1}^{3} \bm{I}^i.\bm{Q}^i.\bm{I}^i.
    \label{hamiltonian}
\end{dmath}
Here, $\bm{S}$, $\bm{I}$ are electron and nuclear spin operators, respectively. $D=3.45$ GHz is the zero field splitting (ZFS) term, $\epsilon$ is the strain term, which has been set to zero for this analysis, $\gamma_e$, $\gamma_n$ are the electron and nuclear gyromagnetic ratio, $\bm{A}$ is the hyperfine interaction tensor, $\bm{Q}$ is the quadrupole interaction tensor. The $z$-axis is the electron spin quantization axis, perpendicular to the hBN plane.
Below, we outline the methodology and key steps involved in the numerical simulation of the composite $\left(V_B^-, ^{14}N\right)$ defect system.
The hyperfine interaction tensor $\bm{A}$ for the first shell $^{14}N$ nuclear spins as outlined in Table \ref{Table I}, is obtained from the parameters reported by Gao \textit{et al.} \cite{Gao2022}. As the first shell $^{14}N$ nuclear spins are arranged symmetrically around the central $V_B^-$ defect, the hyperfine interaction tensors of the other two nuclei can be obtained by rotating the hyperfine tensor of reference nuclei ($^{14}N_1$) about the symmetry axis ($z$-axis). The quadrupole interaction tensor $\bm{Q}$ for the $^{14}N_1$ nuclear spin as given in Table \ref{Table I}, is based on the value reported by Gracheva \textit{et al.} \cite{Gracheva2022}. Using the similar rotational transformation used for hyperfine parameters, we derive the quadrupole tensors for the remaining two $^{14}N$ nuclei.
\renewcommand{\figurename}{FIG.}
We utilize the quantum toolbox in Python (QuTiP) \cite{lambert2024qutip5quantumtoolbox} to construct spin operators for both electronic ($\bm{S}$) and nuclear ($\bm{I}$) subsystems. The ability to handle high-dimensional Hilbert spaces and define tensor products of spin operators made QuTiP a natural choice for these calculations. QuTiP's built-in functions were employed to create the Hamiltonian matrix and compute the eigen-energies and eigen-states of the full spin system. 
Subsequently, we calculate the gradient and curvature associated with the transition energies (see Appendix~\ref{sec:appendixA} for the in-depth procedure), which quantify the first- and second-order effects of magnetic fluctuations in the surrounding bath.
\renewcommand{\figurename}{FIG.}
\begin{figure*}[htp]
    \centering
    \includegraphics[width=1\textwidth, trim={0 2.6cm 0 2.4cm}, clip]{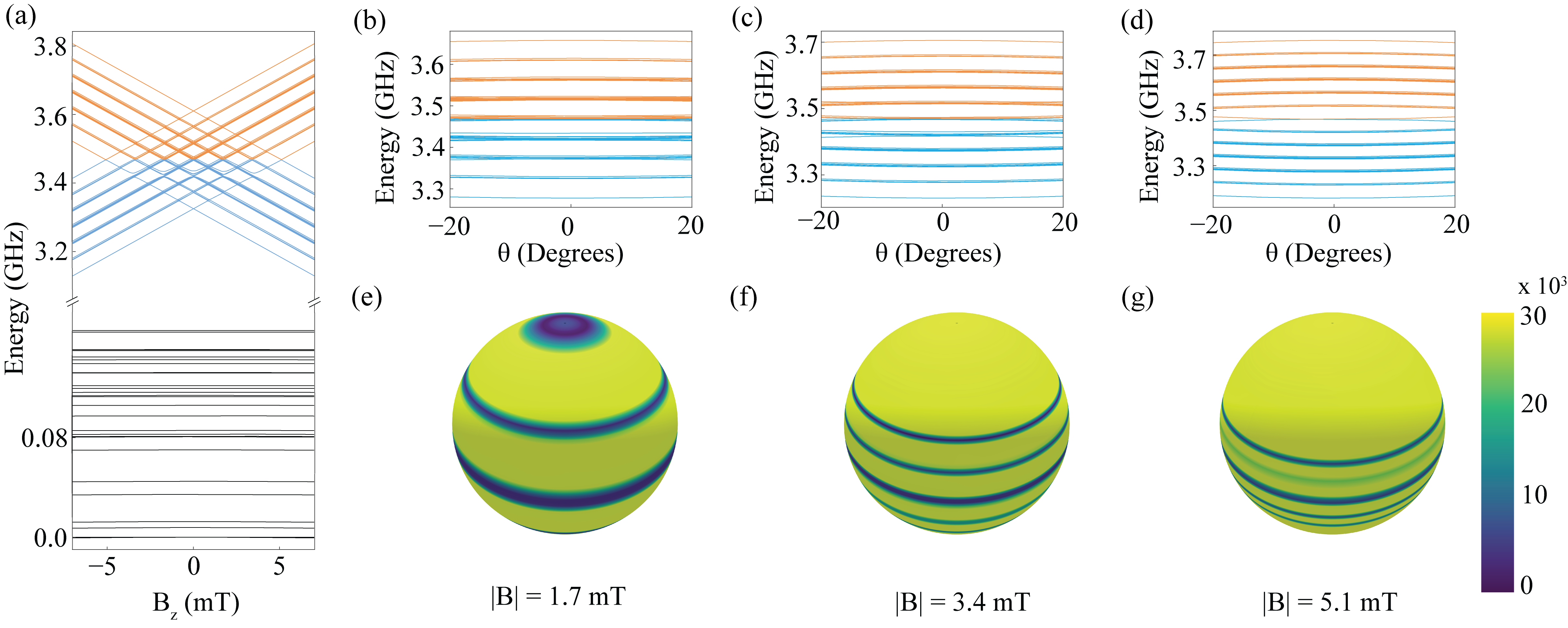}
    \caption{(a) Energy levels of the system as a function of the magnetic field applied parallel to the quantization axis of the electron spin. The black lines correspond to $m_s = 0$, and orange and blue are for $m_s = \pm{1}$ electronic sub-level. These sub-levels split into 27 $(3\times 3\times 3)$ hyperfine levels. The red arrows indicate the chosen magnetic field values. (b), (c) and (d) Energy levels as a function of the polar angle of the applied magnetic field with respect to the quantization axis for 1.7, 3.4, and 5.1 \si{\milli\tesla} magnetic field, respectively. (e), (f) and (g) The magnitude of energy gradient (in MHz/T), as defined in \ref{gradient}, corresponding to the spin transition exhibiting the maximum transition probability at the aforementioned fields. The black arrows represent the $\theta$-range depicted in the figures (b), (c), and (d). The spherical shell radius, in principle, represents the bias magnetic field magnitude \(B_0\) (in \si{\milli\tesla}), however, for visualization purposes, all spheres are shown with the same size, and the radii do not scale with the field strength. The colorbar indicates the magnitude of the energy gradient.}
    \label{fig:2}
\end{figure*}
\section{Results and Discussion}
\begin{figure}
    \includegraphics[width=0.5\textwidth]{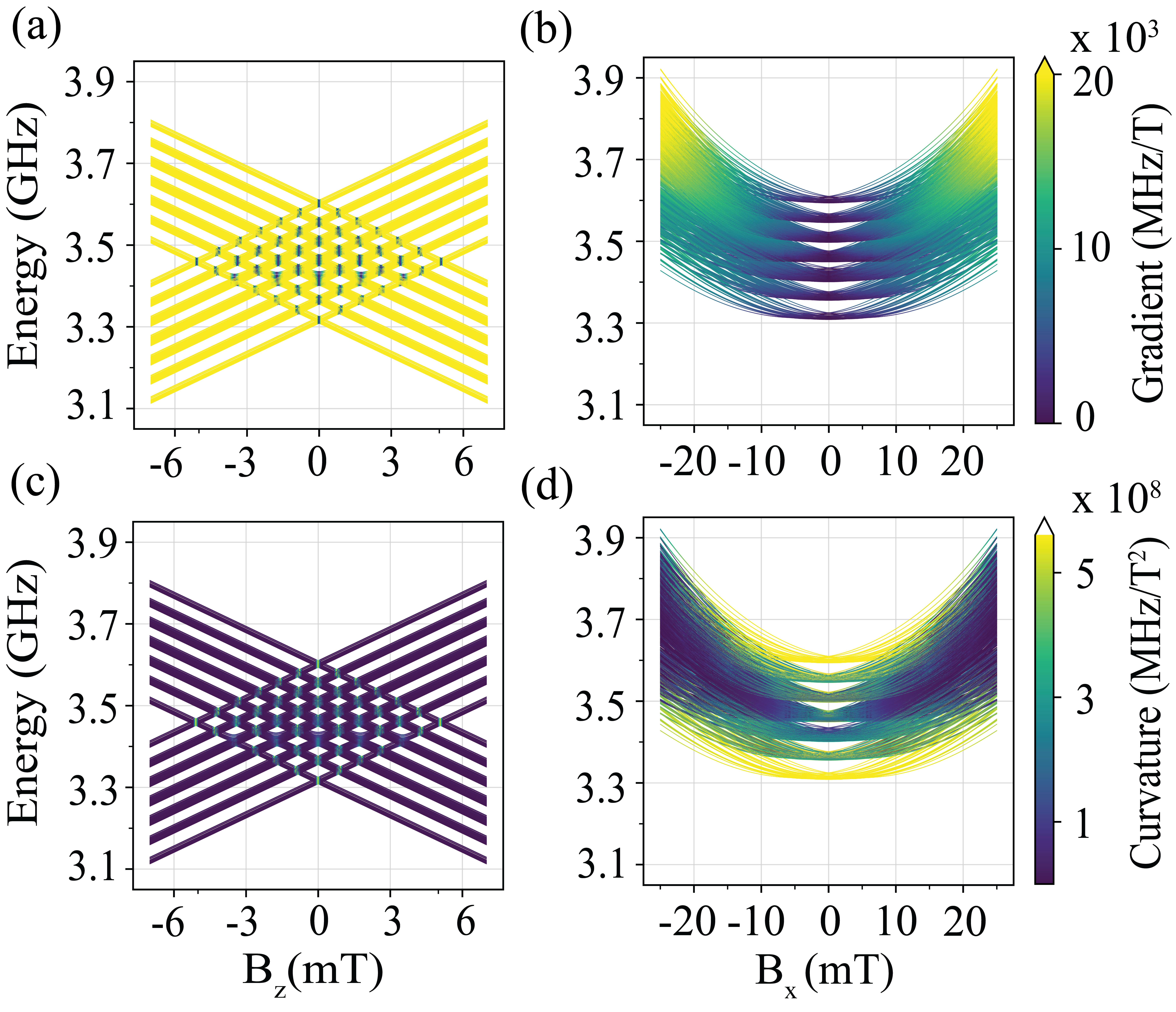}
    \caption{The magnitude of transition energy gradient (as defined in \ref{gradient}) and curvature (as defined in \ref{curvature}) associated with all $(2\times27\times27)$ transitions between $m_s = 0$ and $m_s = \pm{1}$ as a function of the magnetic field applied parallel (a) and (c), perpendicular (b) and (d) to the quantization axis. The color bar corresponds to the magnitude of the gradient and the curvature;  In gradient plots values beyond an arbitrary threshold are whitened for visibility purposes.}
    \label{fig:4}
\end{figure}

Solid-state spin-based sensing protocols typically involve applying a bias magnetic field parallel to the electron spin quantization axis. Here, we perform a comprehensive analysis of the electronic energy level structure of the composite $\left(V_B^-, ^{14}N\right)$ defect as a function of varied bias field amplitude and direction given by:
\begin{equation}
    \bm{B}= B_{0} \bm{\hat{r}}, \quad \bm{\hat{r}} = \left(\sin{\theta}\cos{\varphi},\sin{\theta}\sin{\varphi},\cos{\theta}\right),
\end{equation}
Here, $\theta$ and $\varphi$ are polar and azimuthal angles, respectively, while $B_0$ is the field strength, as shown in Fig. \ref{fig:1}. 
Firstly, Fig. \ref{fig:2}(a) depicts the energy levels of the system as a function of static field strength that is applied along the $z$-axis. 
Here, the primary splitting observed between the $m_s=0$ and $m_s=\pm{1}$ states is due to the zero field interaction term. A smaller splitting between the $m_s=+1$ and $m_s=-1$ sublevels arises from the hyperfine interaction between the electron spin and the three $1^{st}$ shell $^{14}N$ nuclear spins. If the magnetic field is applied along the spin quantization axis, these nuclear spins can be considered equivalent, effectively giving rise to a spin-three system. As a result, the energy sublevels can be grouped into seven distinct hyperfine levels, as seen in Fig. \ref{fig:2}(a).
Additionally, the $m_s=+1$ and $m_s=-1$ levels demonstrate avoided crossings at relatively low fields due to the intricate interplay between Zeeman and hyperfine terms in the system's Hamiltonian; the detailed discussion of this phenomenon is provided in the later part of this section.
Secondly, we investigate the energy level variations as a function of the magnetic field orientation (\(\theta\)) at three distinct anti-crossing points. Figs.~\ref{fig:2}(b)-(d) display the dependence of the electronic energy levels on the polar angle for field strengths of 1.7, 3.4, and 5.1~mT, respectively. In this field regime, the energy levels show minimal dependence on the magnetic field orientation, indicating that they are largely unaffected by local fluctuations.
\renewcommand{\figurename}{FIG.}
\begin{figure*}[htp]
    \centering
    \includegraphics[width=1\textwidth, trim={0 2.45cm 0 2.4cm}, clip]{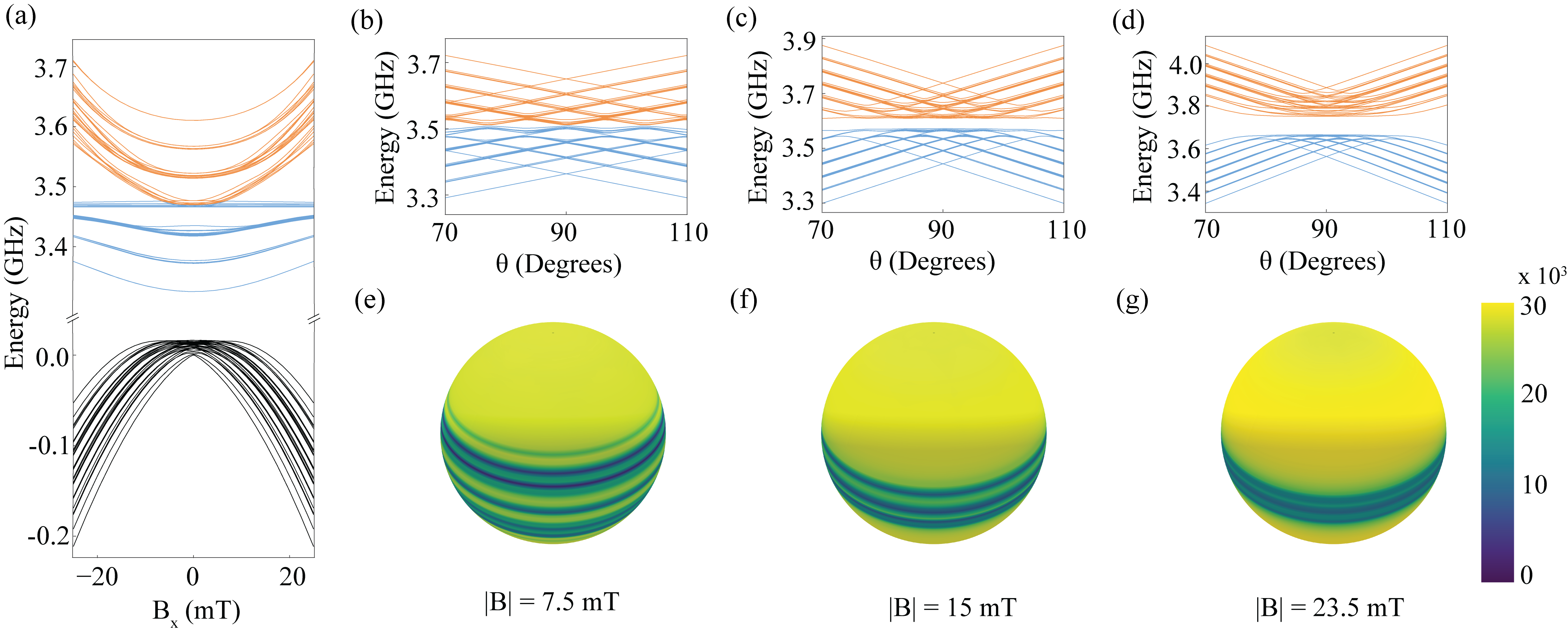}
    \caption{(a) Energy levels as a function of the transverse magnetic field. The black lines correspond to $m_s = 0$, and orange and blue are for $m_s = \pm{1}$ electronic sub-level. The red arrows indicate the chosen magnetic field values. Energy levels as a function of the polar angle of the applied magnetic field with respect to the quantization axis for 7.5, 15, and 23.5 \si{\milli\tesla} magnetic field, respectively. (e), (f) and (g) The magnitude of energy gradient (in MHz/T), as defined in \ref{gradient}, corresponding to the maximum transition probability at the aforementioned fields. The black arrows represent the $\theta$-range depicted in the figures (b), (c), and (d). The spherical shell radius, in principle, represents the bias magnetic field magnitude \(B_0\) (in \si{\milli\tesla}), however, for visualization purposes, all spheres are shown with the same size, meaning the radii do not scale with the field strength. The colorbar indicates the magnitude of the energy gradient.}
    \label{fig:3}
\end{figure*}
To explore the behavior in detail, we vary both the polar angle $\left(\theta\right)$ and azimuthal angles $\left(\varphi\right)$ while keeping the magnitude of the bias field $\left(B_0\right)$  constant. 
The magnitude of gradient of transition energy, $f'$, defined in \ref{gradient}, at each 
$\left(\theta,\varphi\right)$ is mapped to a color scale to visualize the magnetic sensitivity of the transition energy across the sphere. Figs. \ref{fig:2}(e), \ref{fig:2}(f), and \ref{fig:2}(g) illustrate the gradient of the transition energy for the most probable transitions with a transition probability of more than 30\% (see Appendix~\ref{sec:appendixB} for details pertaining to the calculations of transition probabilities).
Our analysis reveals that the gradient at low field strengths is minimal near the poles. As the field strength increases, the gradient gradually shifts towards the equator, forming distinctive rings highly dependent on the bias field. This suggests that beyond this regime of a very low static field ($\leq 5.1$ \si{\milli\tesla}), the low gradient subspaces can be accessed via the application of the field in a transversal direction.
In Fig. \ref{fig:3}(a) we present the ground state energy levels for transverse applied magnetic fields. In this configuration, the primary splitting between $m_s=0$ and $m_s=\pm 1$ states remains dominated by crystal field interactions $(D)$. However, applying the transverse field leads to a significant mixing of the $m_s=+1$ and $m_s=-1$ spin states, significantly modifying the energy level landscape.
We conduct a magnetic field orientation-dependent analysis following the aforementioned approach. In this case, we investigate the energy levels at three distinct magnetic field strengths, namely 7.5, 15, and 23.5 \si{\milli\tesla}, as shown in Figs. \ref{fig:3}(b), \ref{fig:3}(c), and \ref{fig:3}(d), respectively. In this orientation, the $m_s=+1$ and $m_s=-1$ levels undergo anti-crossing, leading to the low $f'$, effectively making the spin states less susceptible to fluctuations in the applied field.
Figs. \ref{fig:3}(e), \ref{fig:3}(f), and \ref{fig:3}(g) depict the energy gradient of a spin transition having highest transition probability within this field regime. 
As anticipated, the lowest gradient values occur near the transverse plane at higher fields. This finding suggests that applying a transverse magnetic field becomes increasingly effective in shielding the spin states from magnetic fluctuations as the field strength is increased, providing enhanced stability and protection for the spin system. Fig.\ref{fig:4}(a) shows the overall gradient of all possible transitions as a function of the applied magnetic field strength oriented parallel to the crystal field axis. At low parallel magnetic fields, the system exhibits notably low gradients at specific field strengths. The sharp dips in the gradient are essentially due to the competing interactions in the Hamiltonian, namely, Zeeman and electron-nuclear. To obtain an intuitive description of these features, we develop an approximate analytical approach based on the perturbation theory. 
\renewcommand{\figurename}{FIG.}
Under a parallel bias magnetic field, the three first-shell $^{14}N$ nuclear spins interact identically to the $V_B^-$ spin. Consequently, the system can be effectively described as an interacting spin-1, spin-3 system. The terms that commute with $S_z^2$ are treated as the unperturbed Hamiltonian $(H^0)$, while the remaining non-commuting terms are taken as the perturbation Hamiltonian $(H^1)$:
\begin{equation}
\begin{aligned}
    H^0 &= DS_{z}^2 + \gamma_e (B_z+b_z) S_z + A_{zz}S_zI_z, \\
    H^1 &= \gamma_e \left(b_x S_x + b_y S_y \right).
\end{aligned}
\end{equation}
Here, $b_x, b_y$ are small fluctuations in the magnetic field components along $x$ and $y$ direction. Applying second order perturbation theory to this Hamiltonian, we obtain the approximate transition energies satisfying the conditions $\Delta m_s = \pm 1, \Delta m_I = 0$, given by (in-depth calculations are provided in the Appendix \ref{sec:appendixC}).
\begin{equation}
    f =  D + m_s\sqrt{\left[\gamma_e\left(B_z + b_z \right) + m_ I A_{zz}\right]^2 + \left[\gamma_e\sqrt{\left(b_{x}^2 + b_{y}^2\right)}\right]^2}.
\end{equation}
The amplitude of the gradient of these transition energies can be obtained using Eq.(\ref{gradient}), Under small magnetic fluctuation condition, one can obtain the minimal gradient at parallel bias fields,
\begin{equation}
    B_z = -\frac{m_{I} A_{zz}}{\gamma_e}.
    \label{dfs_condition}
\end{equation}
For the systems comprising multiple nuclear spin shells, Eq. (\ref{dfs_condition}) will get extended to $B_z = -\sum_{i}\frac{m_{I_i} A_{zz_i}}{\gamma_e},$ where $i$ represents the nuclear spin shell index. If the defect is experiencing out-of-plane interactions due to the presence of multiple hBN layers, the hyperfine coupling $A_{zz}$ needs to be replaced with $\sqrt{A_{zx}^2+A_{zy}^2+A_{zz}^2}$. The matching condition presented in Eq. \ref{dfs_condition} will have to be amended to account for the $\theta$ dependence of the Zeeman term, if the applied field is not aligned with the quantization axis. However, it is apparent that, as the magnetic field strength increases, the matching between Zeeman and hyperfine splittings occurs at larger $\theta$ values.  Consequently, the avoided crossings will shift to larger $\theta$ values for higher magnetic field magnitudes as depicted in Figs. \ref{fig:2} and \ref{fig:3}. Similar phenomena have been observed in nitrogen vacancy centers in diamond \cite{RamaKoteswaraRao2020}.
Furthermore, we analyze the specific scenario when the magnetic field is applied perpendicular to the crystal field axis. Fig. \ref{fig:4}(b) shows that the transition gradients are minimal at zero field, however, their rate of increase as a function of the field strength is dramatically suppressed for the transverse field. A lower gradient indicates a reduced first-order response to external perturbations, implying that the system is less affected by small magnetic fluctuations. In such cases, the second-order effect, characterized by the curvature of the transition energy, becomes more relevant. Therefore, we compute the mean curvature for parallel and transverse field variations (detailed calculations are provided in the Appendix \ref{sec:appendixA}). 
\renewcommand{\figurename}{FIG.}
\begin{figure*}[!]
    \centering
    \includegraphics[width=1\textwidth]{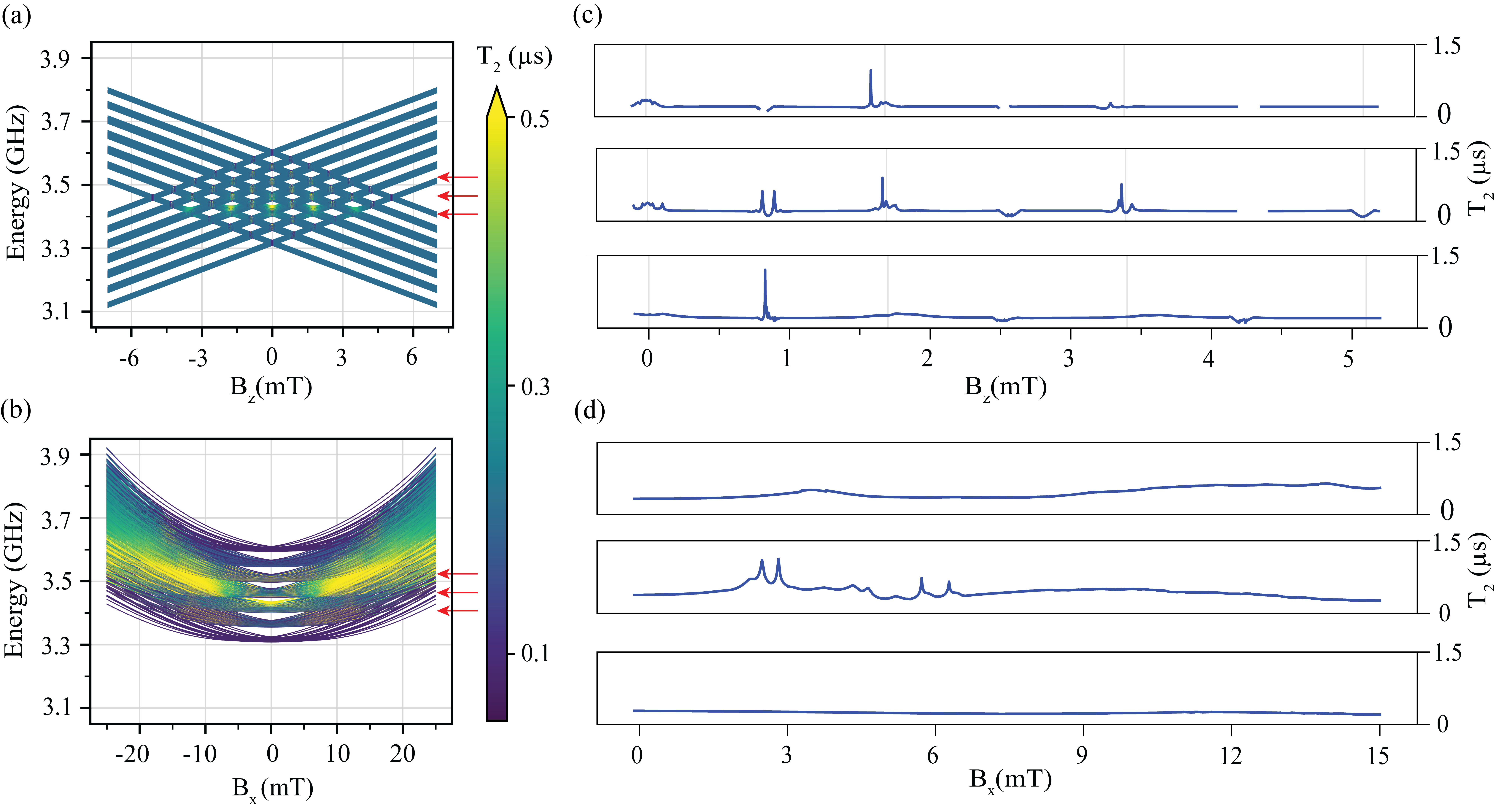}
    \caption{Estimated coherence lifetime as a function of bias magnetic field for (a) parallel field configuration, and (b) transverse field configuration. Here, the upper bound of the colorbar has been restricted to 0.5 $\mu$s for enhancing the visibility of the high $T_2$ regions. Dependence of $T_2$ as a function of the applied field corresponding to select transition energies for (c) parallel field configuration, and (d) transverse field configuration. The red arrows in (a) and (b) indicate the chosen energies, and the gray lines in (c) indicate the magnetic field values that satisfy Eq. \ref{dfs_condition}.}
    \label{fig:5}
\end{figure*}
The curvature (i.e., second order effect due to magnetic fluctuations) of transition energies for fields applied in parallel and transverse orientations are shown in Figs. \ref{fig:4} (c) and \ref{fig:4}(d) respectively. As anticipated from the geometry of the problem, the curvature is larger at the anti-crossing points compared to the other field regions. Besides the anti-crossing points, the curvature increases as a function of the field strength for the parallel field, nonetheless, the magnitude and variations are considerably lower as compared to the transverse field scenario. Moreover, for the transversal field application, although the magnitude of the curvature is frequency dependent, it decreases with magnetic field strength.
Although our analysis does not explicitly facilitate evaluation of the exact $T_2$ values, we implement the following route to illustrate coherence enhancement corresponding to the investigated scenarios. Assuming that $T_2$ is inversely proportional to the frequency broadening caused by the magnetic field perturbations, its dependence on energy gradients and curvatures can be expressed as: 
\begin{equation}
T_2 \approx \frac{1}{\sqrt{\left( f'(B_0) \sigma_B \right)^2 + \frac{1}{2}\left( f''(B_0) \sigma_B^2 \right)^2}}.
\label{T2_expression}
\end{equation} 
where $\sigma_B$ is the standard deviation of the external field noise, assuming the normal distribution (for more details see Appendix \ref{sec:appendixD}). Here, we assume that magnetic noise is the leading source of decoherence, which is the case for $V_B^-$ in hBN due to the nuclear spin dominated bath. Taking into consideration the reported values $T_2$, both experimental as well as those obtained via Cluster Correlation Expansion (CCE) \cite{tarkanyi2025understanding}, we estimate $\sigma_B$ to be 172.6 \si{\micro\tesla} at 15 \si{\milli\tesla}. This allows for the calculation of $T_2$ associated with each of the transition energies at various field configurations using Eq. \ref{T2_expression}. It is important to note that although $\sigma_B$ depends on the magnetic field, its value is effectively constant over the range investigated here \cite{tarkanyi2025understanding}.
The color plots of $T_2$ for the parallel and the perpendicular magnetic field scenarios are presented in Figs. \ref{fig:5}(a) and (b). Moreover, the cross-sections taken at specific transition energies are shown in Figs. \ref{fig:5}(c) and (d). Here, the cross-sections have a width of 10 MHz on each side of the central transition energy. For the parallel fields, we obtain extremely long coherence lifetimes at the peculiar points that satisfy Eq. \ref{dfs_condition}. Additionally, some extra features are observed at frequencies where the approximate analytical solution becomes inadequate to capture the complex energy level structure. The linewidths of the individual $T_2$ peaks are of the order of a few tens of \si{\micro\tesla}, thus requiring a relatively high control over the bias field for the practical realization. On the other hand, the transverse field configurations appear to be much less sensitive to the bias field amplitude, though the $T_2$ enhancement factor is not as high as the one observed for the parallel cases. In an actual experiment, we expect the enhanced $T_2$ features to be less pronounced due to the broadening effects arising from the factors such as interactions with the rest of the bath, bias field inhomogeneities, and the finite bandwidth of the experimental probe pulse. The transition probability of the probed transitions will also influence the outcome of the experiment. Nonetheless, we anticipate noticeably higher coherence lifetimes close to the predicted configurations.
\section{Summary \& Outlook}
Our comprehensive investigation of magnetic field dependent properties of the composite $\left(V_B^-, ^{14}N\right)$ defect uncovers rich physics. In particular, by performing an in-depth numerical simulation of the system's spin Hamiltonian, we identify distinct anti-crossing points, corresponding to specific field strengths and orientations where the energy gradients reach their minimum value. Furthermore, we derive an approximate analytical expression to estimate conditions under which gradient minimization occurs when the magnetic field is oriented along the quantization axis. This model is widely applicable as the low-gradient regions can be evaluated for an arbitrary spin-1 electronic system coupled to an arbitrary nuclear spin. Additionally, we compute curvatures of the transition energies that can play an important role in certain low-gradient regions. Finally, we estimate experimentally observable $T_2$ values and demonstrate that coherence can be improved substantially if specific field configurations are realized.
More broadly, the analysis in this study provides a framework for interpreting magnetic-field-dependent $T_2$ measurements across diverse systems and is expected to encourage further exploration of such coherence-preserving configurations, both experimentally and through advanced numerical methods such as CCE.
Furthermore, our results have significant implications for low-field quantum sensing, where decoherence mitigation remains a critical challenge. Specifically, the identified gradient dips and field-insensitive regions highlight potential pathways for improving spin coherence in $V_B^-$ defects, thus advancing the development of quantum sensing applications such as electometry and thermometry, based on 2D materials.
\section{Acknowledgments}
We are grateful to Professor John J. L. Morton for his valuable insights. S.D. thanks the Indian Institute of Technology, Madras, India, and the Science and Engineering Research Board (SERB Grant No. SRG/2023/000322), India, for start-up funding. S.D. and B.M. acknowledge the use of facilities supported by a grant from the Mphasis F1 Foundation given to the Center for Quantum Information, Communication, and Computing (CQuICC).
\section{Code Availability}
The code for Hamiltonian diagonalization as well as gradient and curvature evaluation is available \href{https://github.com/Basanta-iitm-git/Decoherence_mapping}{here}.
\appendix
\section{Computation of Gradient and Curvature}
\label{sec:appendixA}
The workflow of gradient calculation is as follows: 
\begin{enumerate}
\item At each point on the shell, defined by $\left(\theta,\varphi\right)$ we introduce small fluctuations $\delta \bm{b}$ in magnetic field $\bm{B_{0}}$. Hence, total field strength $\bm{B} = \bm{B_{0}} \pm \delta\bm{b}$.
\item We then find the eigen-energies of the full Hamiltonian and transition energy $\left(f \right)$ for all possible transitions, at each of these points. 
\item Subsequently, we calculate gradient of the transition energy 
$\left(\partial f/\partial b_x, \hspace{.1cm} \partial f/\partial b_y, \hspace{.1cm} \partial f/\partial b_z \right)$ along each direction using central difference method and finally, evaluate the magnitude of the gradient as follows:
\begin{equation}
 f'(B_0) = \sqrt{\left(\frac{\partial f}{\partial b_x}\right)^2 + \left(\frac{\partial f}{\partial b_y}\right)^2 + \left(\frac{\partial f}{\partial b_z}\right)^2}.
\label{gradient}
\end{equation}
\end{enumerate}
The steps for evaluating the curvature are as follows:
\begin{enumerate}
\item At each magnetic field, we introduce small fluctuations directed along six principal axes $(\pm x, \pm y, \pm z)$.
\item We then calculate the eigen-energies of the full Hamiltonian and transition energy $(f)$ for all possible transitions.
\item Subsequently, we calculate first order $(f_x,f_y,f_z)$, second order $(f_{xx},f_{yy},f_{zz})$ derivatives of the transition energy at each magnetic field and finally, evaluate magnitude of the curvature.
\begin{equation}
f''(B_0) = \sqrt{\left(\frac{\partial^2 f}{\partial b_x^2}\right)^2
+ \left(\frac{\partial^2 f}{\partial b_y^2}\right)^2
+ \left(\frac{\partial^2 f}{\partial b_z^2}\right)^2}.
\label{curvature}
\end{equation}
\end{enumerate}
\section{Calculation of Transition Probabilities}
\label{sec:appendixB}
The transition probabilities between spin states in the presence of an oscillating microwave field are computed numerically from the eigenstates of the full spin Hamiltonian. Given initial and final eigenstates \(\ket{\psi_i}\) and \(\ket{\psi_f}\), the transition probability \(P_{i \to f}\) for a microwave excitation polarized perpendicular to the quantization axis is evaluated as the squared matrix element of the electronic spin operators transverse to the field:
\[
P_{i \to f} = \big|\langle \psi_f | S_x | \psi_i \rangle \big|^2 + \big|\langle \psi_f | S_y | \psi_i \rangle \big|^2.
\] 
The transition probabilities for the most probable transitions discussed in Figs.~\ref{fig:2} and \ref{fig:3} are presented in Fig.~\ref{fig:6}.
To further illustrate the relative strength of these transitions, in Fig.~\ref{fig:7} 
These maps highlight the eminence of the selected transitions across varying field strengths and orientations.

\renewcommand{\figurename}{FIG.}
\begin{figure}[ht] 
    \centering
    \includegraphics[width=1\columnwidth]{magfield_tprob.png}
    \caption{Transition probability under (left) parallel and (right) transverse magnetic field. The star-marked points indicate the field values at which detailed analysis of the energy spectra and gradient variations with respect to the magnetic field orientation was performed, which is displayed in Figs. \ref{fig:2}, \ref{fig:3}.}
    \label{fig:6}
\end{figure}

\renewcommand{\figurename}{FIG.}
\begin{figure}[ht] 
    \centering
    \includegraphics[width=1\columnwidth]{Eminense_of_transition.png}
    \caption{Heatmaps illustrating the transition–probability ratios for the two representative transitions discussed in Figs.~2 and 3. The ratio was obtained by dividing the actual transition probability by the probability of the maximally allowed transition within the $m_s=0$ and $m_s=-1$ manifolds. The magnetic-field magnitude $B_{0}$ and polar angle $\theta$ are varied while the azimuthal angle is fixed ($\phi=0$).}
    \label{fig:7}
\end{figure}
\section{Perturbation Analysis}
\label{sec:appendixC}
\setcounter{equation}{0}
Under a parallel bias magnetic field, three $1^{st}$ shell $^{14}N$ nuclear spins are identical to $V_B^-$ spin. Therefore, the system can be considered as an interacting spin-1, spin-3 system. The unperturbed Hamiltonian,
\begin{equation}
    H^0 =  DS_{z}^2 + \left(B_z + b_z\right)S_z + A_{zz}S_zI_z,
\end{equation}
where $B_z$ is the bias magnetic field and $b_z$ is the magnetic field fluctuation along z axis, D and $A_{zz}$ are parallel ZFS and hyperfine constants. The terms which do not commute with $S_z^2$ are considered as part of perturbation Hamiltonian $(H^1)$,
\begin{equation*}
    H^1 =  \gamma_e \left(b_x S_x +b_y S_y \right),
\end{equation*}
\begin{equation}
    H^1 = \gamma_e \left[b_x\left(\frac{S_+ + S_-}{2}\right) + b_y\left(\frac{S_+ - S_-}{2i}\right)\right]. 
\end{equation}
Here, $b_x, b_y$ are small fluctuations in the magnetic field components along $x$ and $y$ direction, $S_+ = S_x + iS_y$ and $S_- = S_x - iS_y$ are the spin-raising and lowering operators.
The unperturbed energy levels corresponding to $H^0$ are given by 
\begin{equation}
    E_{\ket{m_s,m_I}}^0 = Dm_{s}^2 + \gamma_e (B_z+b_z) m_s + A_{zz} m_s m_I
\end{equation}
where $m_s$, electron spin projection quantum number, can take value $\{m_s = 0, \pm1\}$ and $m_I$ is nuclear spin projection quantum number, can take value $\{m_I = -3,..,0,..,3\}$.
As the $\ket{m_s=0,m_I}$ states are degenerate, we apply first-order degenerate perturbation theory, where the corrections to the energies are obtained from the eigenvalues of the perturbation matrix $W^0 = [w]_{7\times7}$. The diagonal and off-diagonal elements of this matrix are,  
\begin{subequations}
\begin{align}
    w^0_{m_I,m_I} & = \braket{0,m_I|H^1|0,m_I} = 0, \\ 
    w^0_{m_I,m_{I}^'} & = \braket{0,m_I|H^1|0,m_{I}^'} = 0.
\end{align}
\end{subequations}
For the $\ket{m_s=\pm 1,m_I}$ states, which are non-degenerate, we use first-order non-degenerate perturbation theory, yielding the energy corrections,  
\begin{equation}
    \braket{\pm1,m_I|H^1|\pm1,m_I} = 0.
\end{equation}
Thus, in both the degenerate and non-degenerate cases, all first-order corrections vanish,  
\begin{equation}
    E_{\ket{m_s,m_I}}^1 = 0.
\end{equation}
Since the degeneracy of the $\ket{m_s=0,m_I}$ states is not lifted at first order, we proceed with second-order degenerate perturbation theory. The effective perturbation matrix is given as,
\begin{equation}
    \Lambda^0 = \left[\lambda\right]_{7\times7},
\end{equation}
where matrix elements are,
\begin{equation}
    \lambda_{ij}^0 = \sum_{n\neq m} 
    \frac{\braket{\varphi_{n,i}|H^1|\psi_m}\braket{\psi_m|H^1|\varphi_{n,j}}}
         {E_{n}^{(0)}-E_{m}^{(0)}} = 0.
\end{equation}
Here, $\varphi_n$ are the degenerate states and $\psi_m$ represent all other states and $E_{n}^{(0)}, E_{m}^{(0)}$ are the zeroth-order energies of the $n^{\mathrm{th}}$ and $m^{\mathrm{th}}$ eigenstate. 
Thus, the second-order energy correction for the $\ket{m_s=0,m_I}$ states also vanishes,
\begin{equation}
    E_{\ket{m_s=0,m_I}}^2 = 0.
\end{equation}
For the $\ket{m_s=\pm 1,m_I}$ states, which are non-degenerate, we apply second-order non-degenerate perturbation theory.  
The general expression for the second-order correction is
\begin{equation}
    E_{\ket{m_s,m_I}}^2 = \sum_{n\neq m} 
    \frac{|{\braket{\psi_m|H^1|\psi_n}}|^2}{E_{n}^{(0)} - E_{m}^{(0)}} ,
\end{equation}
For instance, the correction for the state $\ket{1,3}$ is
\begin{equation}
\begin{aligned}
    E_{\ket{1,3}}^2 
    &= \frac{|\braket{0,3|H^1|1,3}|^2}
    {\gamma_e \left(B_z + b_z\right) + 3A_{zz}} \\[6pt]
    &= \frac{\gamma_{e}^2}{2} \,
       \frac{b_{x}^2 + b_{y}^2}
    {\gamma_e \left(B_z + b_z\right) + 3A_{zz}} .
\end{aligned}
\end{equation}
More generally, the second-order energy correction for $\ket{m_s,m_I}$ can be expressed as
\begin{equation}
    E_{\ket{m_s,m_I}}^2 
    = \frac{m_{s}^2\gamma_{e}^2\left(b_{x}^2 + b_{y}^2\right)}
    {2\left[ m_s \gamma_e \left(B_z + b_z\right) + m_s m_I A_{zz}\right]} .
\end{equation}
The total energy of the system reads:
\begin{equation}
\begin{aligned}
    E & = E^0 + E^1 + E^2 \\
      & = m_{s}^2 D + m_s Y ,
\end{aligned}
\end{equation}
where,
\begin{equation}
    Y = \sqrt{\left[\gamma_e\left(B_z + b_z \right) + m_ I A_{zz}\right]^2 + \left[\gamma_e\sqrt{b_{x}^2 + b_{y}^2}\right]^2}.
\end{equation}
The transition frequency is defined under the condition $\Delta m_s = \pm 1$ and $\Delta m_I = 0$, and is given by:
\begin{equation}
   f = D + m_s \sqrt{\left[\gamma_e\left(B_z + b_z \right) + m_ I A_{zz}\right]^2 + \left[\gamma_e\sqrt{b_{x}^2 + b_{y}^2}\right]^2 }
\end{equation}
The partial derivatives of the transition frequency with respect to $b_x$, $b_y$, and $b_z$ are as follows:
\begin{equation}
    \frac{\partial f}{\partial b_x} =  \frac{ m_{s}\gamma_{e}^2 b_x}{ \sqrt{\left[\gamma_e\left(B_z + b_z \right) + m_ I A_{zz}\right]^2 +\left[\gamma_e\sqrt{b_{x}^2 + b_{y}^2}\right]^2}},
\end{equation}
\begin{equation}
    \frac{\partial f}{\partial b_y} =  \frac{m_{s}\gamma_{e}^2 b_y}{ \sqrt{\left[\gamma_e\left(B_z + b_z \right) + m_ I A_{zz}\right]^2 + \left[\gamma_e\sqrt{b_{x}^2 + b_{y}^2}\right]^2}},
\end{equation}
and
\begin{equation}
    \frac{\partial f}{\partial b_z} =  \frac{m_s\gamma_e \left[\gamma_e\left(B_z + b_z \right) + m_ I A_{zz}\right]}{\sqrt{\left[\gamma_e\left(B_z + b_z \right) + m_ I A_{zz}\right]^2 + \left[\gamma_e\sqrt{b_{x}^2 + b_{y}^2}\right]^2}}.
\end{equation}
The gradient of the transition energies is then given by:
\begin{equation}
    \left|\frac{df}{db}\right| =  \sqrt{\left(\frac{\partial f}{\partial b_x}\right)^2+\left(\frac{\partial f}{\partial b_y}\right)^2+\left(\frac{\partial f}{\partial b_z}\right)^2}.
\end{equation}
The gradient is minimized when:
\begin{equation}
    \left|\frac{df}{db}\right| = 0, 
    \implies 
    \gamma_e B_z + m_I A_{zz} = 0.
\end{equation}
Under the condition of small perturbation $i.e.,$ $\gamma_eB_z, A_{zz} \gg \gamma_e b_x,\gamma_e b_y, \gamma_e b_z$, the bias magnetic field $B_z$ satisfies:
\begin{equation}
    B_z = -\frac{m_I A_{zz}}{\gamma_e}.
\end{equation}
\section{Detailed Derivation of $T_2$ Formulae}
\label{sec:appendixD}
Consider the transition energy $f(\bm{B})$ as a function of the magnetic field $\bm{B}$, which fluctuates around a mean field value $\bm{B_{0}}$. Expanding $f(\bm{B})$ using a Taylor series about $\bm{B_{0}}$ gives:
\begin{equation}
    \begin{aligned}
     f(\bm{B}) &\approx f(\bm{B}_0) + \nabla f(\bm{B}_0) \cdot (\bm{B} - \bm{B}_0) \\
    &\quad + \frac{1}{2} (\bm{B} - \bm{B}_0)^T \, H_f(\bm{B}_0) (\bm{B} - \bm{B}_0) + ...
    \end{aligned}
\end{equation}
where $\nabla f(\bm{B}_0)$ is gradient of \textit{f} at $\bm{B_{0}}$ and $H_f(\bm{B_{0}})$ is Hessian matrix at $\bm{B_{0}}$. Assuming the magnetic field fluctuations $\delta\bm{b} = \bm{B} - \bm{B_{0}}$ are described by a Gaussian distribution with zero mean value and standard deviation $\bm{\sigma_{B}}$. The variance of the energy fluctuations $\sigma_f^2 = \mathrm{Var}(f)$ can be approximated by truncating higher order terms:
\begin{equation}
\begin{split}
\sigma_f^2 &= \mathrm{Var} \left( f(\bm{B}_0) + \nabla f(\bm{B_{0}}) \cdot \delta\bm{b} + \frac{1}{2} \delta \bm{b}^T H_f(\bm{B_{0}}) \delta\bm{b}\right) \\
&= 0 + \left[ \left( \frac{\partial f}{\partial b_x} \right)^2 + \left( \frac{\partial f}{\partial b_y} \right)^2 + \left( \frac{\partial f}{\partial b_z} \right)^2 \right] \, \sigma_B^2 \\ 
&+ \frac{1}{2} \left[ 
    \left(\frac{\partial^2 f}{\partial b_x^2}\right)^2 +
    \left(\frac{\partial^2 f}{\partial b_y^2}\right)^2 +
    \left(\frac{\partial^2 f}{\partial b_z^2}\right)^2
    \right] \sigma_B^4
\end{split}
\end{equation}
where we have used the fact that for a Gaussian variable $\delta b$, $\mathrm{Var}(\delta b) = \sigma_B^2$ and $\mathrm{Var}((\delta b)^2) = 2 \sigma_B^4$.

The coherence time $T_2$ can then be approximated as:
\begin{equation}
T_2 \approx \frac{1}{\sigma_f} = \frac{1}{\sqrt{\left[ f'(B_0) \sigma_B \right]^2 + \frac{1}{2} \left[ f''(B_0) \sigma_B^2 \right]^2}}.
\end{equation}
Here, 
\[f'(B_0) = \sqrt{\left( \frac{\partial f}{\partial b_x} \right)^2 + \left( \frac{\partial f}{\partial b_y} \right)^2 + \left( \frac{\partial f}{\partial b_z} \right)^2}\]
\[f''(B_0) = \sqrt{\left(\frac{\partial^2 f}{\partial b_x^2}\right)^2 + \left(\frac{\partial^2 f}{\partial b_y^2}\right)^2 +\left(\frac{\partial^2 f}{\partial b_z^2}\right)^2}\]
This expression (Eq.\ref{T2_expression} in the main text) quantifies how both linear (gradient) and quadratic (curvature) sensitivities of the transition energy to magnetic field fluctuations influence the spin coherence time.
\vspace{0.2in}

\bibliographystyle{apsrev4-2}
\bibliography{dfs_ref_1.bib}

\end{document}